\documentclass[pdflatex,sn-mathphys-num]{sn-jnl}% Math and Physical Sciences Numbered Reference Style
\usepackage{anyfontsize}
\usepackage{graphicx}%
\usepackage{multirow}%
\usepackage{amsmath,amssymb,amsfonts}%
\usepackage{amsthm}%
\usepackage{mathrsfs}%
\usepackage[title]{appendix}%
\usepackage{xcolor}%
\usepackage{textcomp}%
\usepackage{manyfoot}%
\usepackage{booktabs}%
\usepackage{algorithmicx}%
\usepackage{algorithm}%
\usepackage{algpseudocode}%
\usepackage{listings}%
\usepackage{hyperref}
\usepackage{makecell}
\usepackage{diagbox}
\usepackage{subcaption}
\theoremstyle{thmstyleone}%
\theoremstyle{thmstyletwo}%

\theoremstyle{thmstylethree}%

\begin{document}

\title[ODEONN: A Digital ODE Solver Architecture for Oscillatory Neural Networks]{ODEONN: A Digital ODE Solver Architecture for Oscillatory Neural Networks}


\author[1]{\fnm{Bram F.} \sur{Haverkort}}%\email{iauthor@gmail.com}

\author*[1]{\fnm{Aida} \sur{Todri-Sanial}}\email{a.todri.sanial@tue.nl}

\affil[1]{\orgdiv{NanoComputing Research Lab, Integrated Circuits Group, Department of Electrical Engineering}, \orgname{Eindhoven University of Technology}, \orgaddress{\street{Groene Loper 3}, \city{Eindhoven}, \postcode{5612 AE}, \state{Noord-Brabant}, \country{The Netherlands}}}


% \abstract{The abstract serves both as a general introduction to the topic and as a brief, non-technical summary of the main results and their implications. Authors are advised to check the author instructions for the journal they are submitting to for word limits and if structural elements like subheadings, citations, or equations are permitted.}

\abstract{
    Oscillatory Neural Networks (ONNs) are an alternative computing paradigm for AI and combinatorial optimization problems.
    However, digital architectures are often designed for specific applications of ONNs.
    This work introduces a modular and scalable architecture called ODEONN that is generic to multiple applications of ONNs, and to the best of our knowledge, is the first fully digital ONN to also support complex-valued coupling.
    Additionally, an approximation of the sine function is introduced that uses half of the hardware resources compared to standard methods.
    The performance of ODEONN is compared with a full-precision software simulation, where a performance degradation of less than $2\%$ is shown.
    Therefore, we conclude that the fixed-point quantization and the approximated waveform affect the accuracy of computation by only a small amount.
    Furthermore, ODEONN shows a 45$\times$ reduction in energy-delay product over the software simulation running on conventional hardware.
}




\maketitle

\section{Introduction} \label{sec:introduction}

Oscillatory Neural Networks (ONNs) are networks of coupled oscillators \cite{todri-sanial_computing_2024}, which have been shown perform gradient descent in an abstract energy landscape defined by their relative coupling strengths \cite{hoppensteadt_pattern_2000}.
By carefully choosing the coupling strengths, problems can be embedded in the ONN using Ising formulations \cite{lucas_ising_2014}.
The relative phase differences between the oscillators at equilibrium will convey information about the solution.
Using this core principle, many algorithms have been developed; from AI tasks, such as associative memory and classification \cite{cai_oscnet_2025, sabo_classonn_2024, gower_how_2025}, to graph-based combinatorial optimization problems such as max-cut and traveling salesperson \cite{vadlamani_combinatorial_2024, landge_n-oscillator_2020, gonul_gpu-accelerated_2025, su_roc-spin_2024, wang_solving_2021}.

ONNs are often implemented using analog circuits or spintronics \cite{soni_spinonn_2025, cilasun_coupled-oscillator-based_2025, su_roc-spin_2024}, as this naturally fits the paradigm.
However, concessions have to be made in these implementations by limiting either the overall network size or sparsifying the connectivity between oscillators.
Both options have advantages and disadvantages.
A smaller fully-connected network allows for easy problem embedding, but limits the scale of problems that can be solved.
In contrast, a larger network with a sparse topology, will allow for the embedding of larger scale problems, but the embedding process itself can become complicated \cite{li_improved_2025}.

Digital ONN architectures have come forward as an alternative to analog ONNs that allow for designs that scale easier, while maintaining fully connected networks \cite{bashar_fpga-based_2024, sreedhara_digital_2023, abernot_digital_2021, delacour_oscillatory_2021, Gonul2026}.
However, many of these implementations only allow for binary phase measurements, so either 0 or $\pi$ radians, or lack precision by limiting the phase to a small number of discrete steps.
Others have introduced ordinary differential equation (ODE) Solver based digital ONN architectures, but have targeted a specific application \cite{bashar_fpga-based_2024} or use simplified waveforms, such as square waves \cite{liu_efficient_2024, DeepGoru2026}.
In this work, we introduce ODEONN, a digital ODE solver architecture implemented on a Field Programmable Gate Array (FPGA) dedicated to solving the ODE that describes ONN behavior, without targeting a specific application for ONNs and while maintaining a smooth waveform function, ideally a pure sine wave.

ODEONN compares favorably with a software-based ODE solver running on a graphics processing unit (GPU), while utilizing as little resources as possible.
To our knowledge, this paper is also the first to introduce a versatile digital ONN architecture that includes support for complex coupling.

This paper is structured as follows:
Section \ref{sec:background} gives the required background knowledge on ONNs and ODE Solvers.
In Section \ref{sec:architecture}, the architecture and its sub-components will be explained.
Sections \ref{sec:methods} and \ref{sec:results} will go into the benchmarking methodology and the results thereof, respectively.
Finally, this paper will be concluded in Sections \ref{sec:conclusion} and \ref{sec:discussion}.

\section{Background} \label{sec:background}

To describe the dynamics of an ONN, the Kuramoto model is used \cite{Kuramoto_1984} :
\begin{equation}
    \label{eq:background:kuramoto}
    \dot\phi_i = \omega_i + \sum_jC_{i,j}\sin(\phi_j-\phi_i),
\end{equation}
where $\omega_i$ is the frequency and $\phi_i$ is the phase of the $i$'th oscillator.
$C_{i,j}$ is the coupling coefficient between oscillators $i$ and $j$.
If one takes $C_{i,j}$ to be a complex number, it can be expanded such that 
\begin{equation}
    \label{eq:background:coupling}
    C_{i,j} = W_{i,j}e^{\text{i}\delta_{i,j}},
\end{equation}
where $W_{i,j}$ is the coupling strength between oscillators $i$ and $j$, and $\delta_{i,j}$ is the phase angle of the coupling between oscillator $i$ and $j$.
Using complex couplings, phase angles between oscillators other than $0$ and $\pi$ can be set.
In this work, a uniform fixed frequency of all oscillators is assumed, therefore $\omega_i$ is an integration constant and can be neglected.
Additionally, one can use the identity given in Eq. \eqref{eq:background:coupling} and the complex exponential notation of a sine to absorb the $\delta_{i,j}$ term into the $\sin(\phi_j - \phi_i)$ term and use the $W_{i,j}$ term as a scalar weight.
Thus, the following simplified equation, which is implemented in this work, is given:
\begin{equation}
    \label{eq:background:kuramoto_simple}
    \dot\phi_i = \sum_jW_{i,j}\sin(\phi_j-\phi_i-\delta_{i,j}).
\end{equation}

\section{Architecture} \label{sec:architecture}

\subsection{Architecture Overview} \label{sec:architecture:overall}

\begin{figure}
    \centering
    \includegraphics[width=\linewidth]{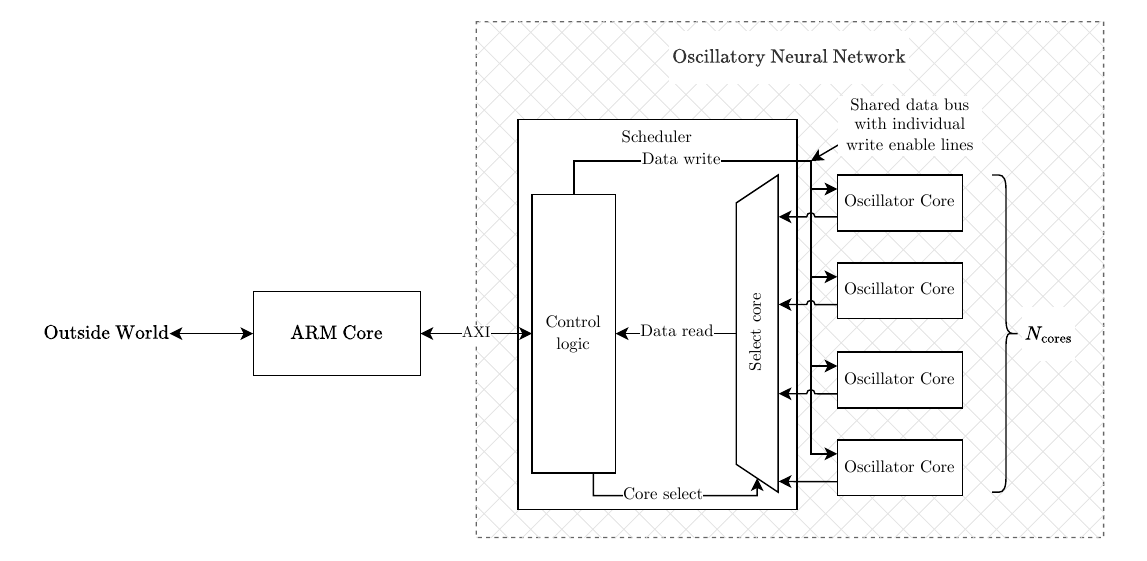}
    \caption{Block diagram of the ODEONN architecture. An external interface is used to program a scheduler, which in turn controls the computation in the oscillator cores and facilitates the communication between them. The scheduler selects cores one by one to read data from it, and transmits it to the other cores via a shared databus. The scheduler is also connected to an external device, in this example an ARM Core, through an Advanced eXtensible Interface (AXI) interface.}
    \label{fig:architecture:overall}
    \label{fig:architecture:scheduler}
\end{figure}

Fig. \ref{fig:architecture:overall} shows a generalized layout of the ODEONN architecture, comprising of a global scheduler and $N_\text{cores}$ number of oscillator cores, which can be determined at time of hardware synthesis.
The scheduler facilitates the communication between each oscillator core and the external communication through an external interface, in the case of the chosen FPGA board, a PYNQ-Z2, an Advanced eXtensible Interface (AXI) interface to an ARM processor.
This interface is used to upload the coupling strength and coupling phase angles, which determine the network configuration, and to set the desired number of cycles to compute.
After the computation is complete, the same AXI interface is used to download the final phase state.
The phase evolution can also be streamed during computation using an AXI-stream interface.
The oscillator cores are the main functional units that replicate the behavior of an oscillator in an ONN.

\subsection{Scheduler} \label{sec:architecture:scheduler}

The scheduler block, as shown in Fig. \ref{fig:architecture:scheduler}, contains the control logic of the architecture.
It facilitates the communication between each of the Oscillator Cores and can be interfaced with via AXI.
The interface utilizes a standard AXI interface and is connected to an ARM CPU core, as provided by the PYNQ-Z2 hardware.
Communication between the Oscillator Cores is achieved by using a parallel bus interface for writing data, and a multiplexed system for reading data.
The data bus and address bus for writing data are shared across all Oscillator Cores, but the scheduler has a write enable line per Oscillator Core.
This setup allows data to be written to individual cores with a single data bus.
The enable lines can also be activated simultaneously to write the same value to all Oscillator Cores in only one cycle, which is used in the control algorithm.
Additionally, the scheduler allows each of the Block Random Access Memory (BRAM) units present in the oscillator cores to be accessed via the AXI interface, which enables the weight matrices to be programmed.
To read data, a multiplexer is used to select individual data lines of each Oscillator Core.
Finally, there are three control signals, one to start the computation, one to set the total number of integration steps to compute, and one that raises a flag once this iteration count has been reached.
To track the iteration count, there is a simple counter that increments each time the phases of all Oscillator Cores are updated.
Once the network has been programmed and the control flag to start the computation has been raised, the scheduler will perform Algorithm \ref{alg:scheduler} for the set number of iterations.
\begin{algorithm}
    \caption{Algorithm for scheduling data transfer between Oscillator Cores}
    \label{alg:scheduler}
    \begin{algorithmic}
        \For {$N_\text{iterations}$}
            \For {$i = 1$ to $N_\text{cores}$}
                    \State Obtain the phase $\phi_i$ from core $i$
                    \State Transmit $\phi_i$ to each other core
            \EndFor
            \State Wait for the cores to complete this iteration
        \EndFor
        \State Raise the flag to indicate the end of computation
    \end{algorithmic}
\end{algorithm}

\subsection{Oscillator Core} \label{sec:architecture:oscillator_core}

\begin{figure}
    \centering
    \includegraphics[width=\linewidth]{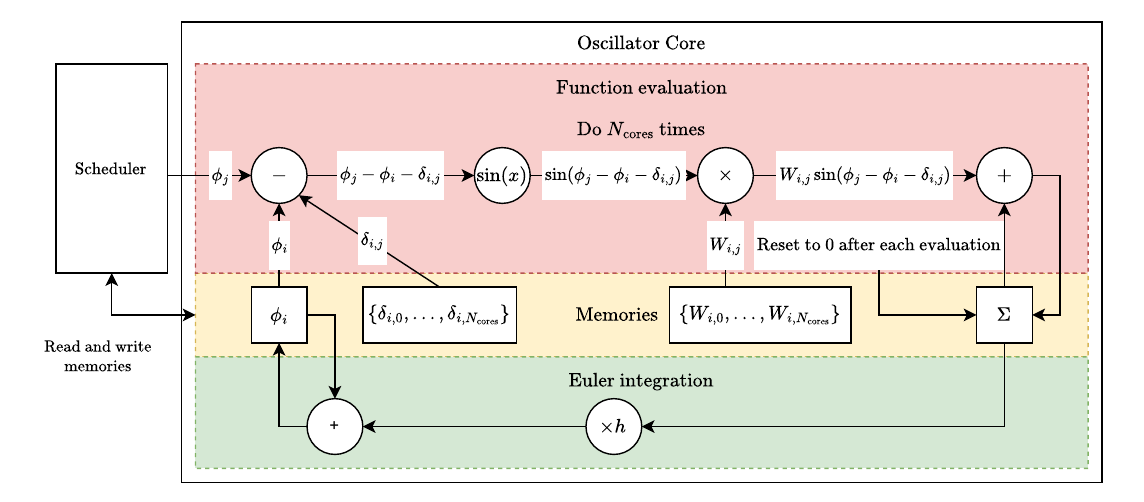}
    \caption{Block level overview of the oscillator core. It is divided into three sections: the function evaluation, the memories, and the Euler integration.}
    \label{fig:architecture:oscillator_core}
\end{figure}

A block level view of the oscillator core is shown in Fig. \ref{fig:architecture:oscillator_core}.
The oscillator core is divided into three main parts: the function evaluation, the memories, and the Euler integration.
The function evaluation section evaluates the value of Eq. \ref{eq:background:kuramoto_simple} at each iteration of the integrator.
A pipelined computation is used for the function evaluation section, whose stages are shown in Fig. \ref{fig:architecture:oscillator_core} at the top in red.
The computation proceeds as follows: first $\phi_j$ and index $j$ is obtained from the scheduler.
The next step is to compute the phase sum $\phi_j - \phi_i - \delta_{i,j}$.
This value is then sent to the waveform evaluation unit, which computes the sine of the given phase value.
The exact details of the functionality of this unit is explained later in Section \ref{sec:architecture:sine_approximation}.
Next, the evaluated sine function is multiplied by the coupling strength $W_{i,j}$.
These values are then accumulated for $N_\text{cores}$ cycles.
After a complete function evaluation, the value is transferred to the integration stage, and the accumulated value is reset to $0$ in preparation for the next iteration.

The integration section uses a first order forward Euler integration method \cite{Euler1768}, namely
\begin{equation}
    \label{eq:architecture:euler}
    \phi_{t+1} = \phi_t + h\dot\phi_t,
\end{equation}
where $h$ is an integration step size constant.
The hardware cost of such an integration scheme can be much lower, since only a single multiplier is used, or when $h$ is an exact power of $2$, a bitshift can replace the multiplier.
Previous similar architectures have also used Euler integration schemes \cite{bashar_fpga-based_2024, liu_efficient_2024, sreedhara_digital_2023, Gonul2026, DeepGoru2026} and have shown promising results.
Other integration schemes are possible, but can require evaluating $\dot{\phi_t}$ multiple times, for example a Runge-Kutta method, or solving an algebraic equation, for example an implicit integration method, both of which can be computationally expensive.
For these reasons, the Euler integration method is chosen.

There are four memory units: the phase memory, the coupling strength memory, the coupling phase memory, and the accumulator.
The two coupling memories are implemented in BRAM units, due to their large size and the fact that they are accessed only serially.
The BRAM unit size $N_{b, \text{BRAM}}$ depends on the number oscillator cores that are implemented $N_\text{cores}$ and the number of bits per coupling strength and coupling phase, $N_{b, W_{i,j}}$ and $N_{b, \delta_{i,j}}$ by
\begin{equation}
    N_{b, \text{BRAM}} = N_\text{cores}\left( N_{b, W_{i,j}} + N_{b, \delta_{i,j}} \right).
\end{equation}
The other two memories are implemented as registers, since they are singular values.

\subsection{Quantization} \label{sec:architecture:quantization}

The main quantity on which the computation is based, the phase $\phi$ has a bounded value $[0, 2\pi)$.
Due to the periodicity of the sine function, any overflow or underflow of this value results in the value wrapping around without impacting the computation.
Therefore, in the computation hardware any overflow or underflow of $\phi$ is allowed.
This bounded range lends itself naturally to fixed-point computation;
The exact minimum and maximum values that the quantity takes on are known beforehand.
Hence, fixed-point quantization is used for ODEONN.
The full dynamic range of a floating-point system would be redundant.
The number of bits for the phase was chosen to be 16, since this fits well into the existing 32-bit AXI interface, and is still within the bounds of the digital signal processing (DSP) hardware in the FPGA, which allows for multiplications up to 18 by 25 bits.
As a result, the range $[0, 2\pi)$ is mapped to $[0, 2^{16})$.
$\pi$ is factored out and is accounted for in the $\sin$ evaluation unit, which is explained in Section \ref{sec:architecture:sine_approximation}.
The number of bits of the evaluated sine is chosen to be the same as the number of bits for $\phi$, for the same reasons.

The quantization of the weight quantities $W_{i,j}$ and $\delta_{i,j}$ is arbitrary, and both are set to 8 bits.% to fit into the BRAM data bus.
This allows the weight phase angle to be set to an accuracy of $\sim0.025$ radians or $\sim1.41$\textdegree.
In later sections the quantized quantities will be denoted using $\widehat{W_{i,j}}$ and $\widehat{\delta_{i,j}}$ for brevity.
As a quick reference, the quantization level of each variable is given in Table \ref{tab:quantization:levels}.

The local truncation error of Euler's method is proportional to $h^2$, while the global truncation error is proportional to $h$.
To obtain a global error of less that $1\%$, which is an arbitrary choice, when talking in terms of phase error, $h \leq 0.01\cdot2\pi \lesssim 0.06$.
$h$ should be ideally be an exact power of two to save computation hardware as discussed earlier, therefore $h \leq 2^{\lceil\log_2(0.06)\rceil} = 2^{-4}$.
For extra margin, $h = 2^{-8} \approx 0.0039$ was used.

\begin{table}
    \captionsetup{width=\textwidth}
    \caption{Quantization levels of the variables present in Eq. \eqref{eq:background:kuramoto_simple}}
    \centering
    \begin{tabular}{|c|c|} \hline
        Variable            & Number of bits    \\ \hline
        $\phi$              & 16                \\
        $\sin$ amplitude    & 16                \\ 
        $W_{i,j}$           & 8                 \\
        $\delta_{i,j}$      & 8                 \\ \hline
    \end{tabular}
    \label{tab:quantization:levels}
\end{table}

\subsection{Sine Approximation} \label{sec:architecture:sine_approximation}

As part of a function evaluation for each oscillatory integrator unit, a sine function needs to be computed.
Typically, in digital hardware either a look-up table or a CORDIC algorithm \cite{Volder1959} is used to compute sine evaluations.
However, look-up tables are not feasible to scale up for larger number of phase bits; for example, a 16-bit valued sine lookup table would require $2^{16} = 65536$ individual table entries, or $2^{14} = 16384$ when storing a quarter wave, which is enough to fully represent a sine wave, by flipping and reflecting the quarter wave as needed. 
At 16 bits per sample, this would already require about 4096 FPGA look-up-tables (LUTs), when used as 64-bit read only memories, which would already utilize over $7.5\%$ of the total LUTs available on the chosen FPGA. 

Alternatively, one could use the aforementioned CORDIC algorithm.
CORDIC, or hybrid versions of it, has found use in accelerators similar to the one discussed in this work \cite{bashar_fpga-based_2024}, however, it remains expensive to implement many CORDIC units in parallel.

\begin{figure}
    \centering
    \begin{subfigure}{0.47\textwidth}
        \centering
        \includegraphics[width=\linewidth]{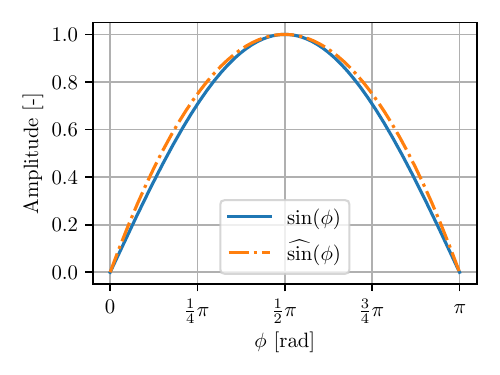}
        \caption{}
        \label{fig:arch_sine_approx:amplitude}
    \end{subfigure}
    \hfill
    \begin{subfigure}{0.475\textwidth}
        \centering
        \includegraphics[width=\linewidth]{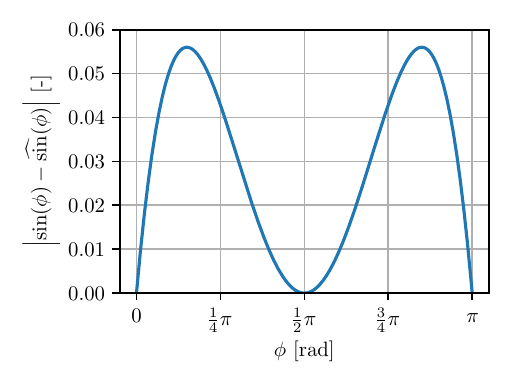}
        \caption{}
        \label{fig:arch_sine_approx:error}
    \end{subfigure}
    \caption{The approximation of the sine function $\widehat{\sin}(\phi)$.\\(a) First half period of a sine function $\sin(\phi)$ and the quadratic approximation $\widehat{\sin}(\phi)$. The second half period is not shown for clarity as it is symmetric regardless.\\(b) Absolute error of the quadratic approximation for the first half period. The second half period is not shown for clarity as it is symmetric regardless.}
    \label{fig:arch_sine_approx}
\end{figure}

It can be argued that for the use case of emulating the behavior of an oscillatory neural network, an exact evaluation of the sine function is not required, and other waveforms have successfully been used \cite{liu_efficient_2024}.
Therefore, the following quadratic approximation of the sine function is proposed.
\begin{equation}
    \label{eq:arch_sine_approx:quadratic_approximation}
    \widehat{\sin}(\phi) := 
    \left\{
    \renewcommand{\arraystretch}{1.5}
    \begin{array}{cl}
        \frac{4\phi(\pi-\phi)}{\pi^2}, & \text{if } 0 \leq \phi < \pi \\
        -\frac{4(\phi-\pi)(2\pi-\phi)}{\pi^2}, & \text{if } \pi \leq \phi < 2\pi 
    \end{array}
    \right.
\end{equation}
The main feature of this approximation is that it can be implemented using a single multiplication with two non-constant values, $\phi$ and $\pi - \phi$.
The other multiplication and division can be implemented using simple bit shifts, because each constant is an exact power of two.
Recall that the phase is quantized such that the range $[0, 2\pi)$ is mapped to $[0, 2^{16})$, which means that $\pi := 2^{15}$.
Additionally, the zero crossings, minima, and maxima all lie at the exact same values as a true sine function.
The maximum absolute error of the approximation given by equation \eqref{eq:arch_sine_approx:quadratic_approximation} is approximately $0.056$, but most importantly the zeros and maxima are exact.
For illustration, Figs. \ref{fig:arch_sine_approx:amplitude} and \ref{fig:arch_sine_approx:error} show the amplitude and absolute error of the given quadratic error and the exact sine function for the first half period, respectively.
The second half period is a shifted and negated version, and therefore has the same error values and is excluded from the figure for readability.
We show that this error is small enough to not affect the outcome of the computation by much, which is verified through several benchmarks in Sections \ref{sec:methods} and \ref{sec:results}.

To compare the resource usage of these three approaches, each of them has been synthesized and implemented for 16-bit sine evaluation, the results of which are shown in Table \ref{tab:arch_sine_approx:resource_usage}.
As shown in Table \ref{tab:arch_sine_approx:resource_usage}, a CORDIC algorithm uses $1.37\%$ of the FPGA LUT resources, making it impossible to implement more than $72$ units, while disregarding the resources needed for the rest of the system.
In contrast, the approximated sine wave allows for the implementation of 128 units, either with or without a DSP unit, while not over-utilizing the FPGA resources and leaving room for other logic.
\begin{table}
    \captionsetup{width=\textwidth}
    \caption{FPGA resource usage of different 16-bit sine implementations.}
    \centering
    \begin{tabular}{|c|c|c|c|} \hline
        Sine Algorithm & LUTs & FFs & DSPs \\ \hline
        \multicolumn{4}{|c|}{Single unit} \\ \hline
        Lookup Table & 5008 \textbf{(9.41\%)} & 16 \textbf{(0.02\%)} & 0 \\ 
        CORDIC & 731 \textbf{(1.37\%)} & 704 \textbf{(0.66\%)} & 0 \\
        Eq. \eqref{eq:arch_sine_approx:quadratic_approximation} & 58 \textbf{(0.11\%)} & 48 \textbf{(0.04\%)} & 1 \textbf{(0.45\%)} \\
        Eq. \eqref{eq:arch_sine_approx:quadratic_approximation} (No DSP) & 291 \textbf{(0.55\%)} & 79 \textbf{(0.07\%)} & 0 \\ \hline
        \multicolumn{4}{|c|}{Scaled to 128 units} \\ \hline
        Lookup Table & 641k \textbf{\textcolor{red}{(1204\%)}} & 2048 \textbf{(1.92\%)} & 0 \\ 
        CORDIC & 93.6k \textbf{\textcolor{red}{(176\%)}} & 90.1k \textbf{(84.7\%)} & 0 \\
        Eq. \eqref{eq:arch_sine_approx:quadratic_approximation} & 7424 \textbf{(14.0\%)} & 6144 \textbf{(5.12\%)} & 128 \textbf{(58.2\%)} \\
        Eq. \eqref{eq:arch_sine_approx:quadratic_approximation} (No DSP) & 37.2k \textbf{(70.0\%)} & 10.1k \textbf{(8.96\%)} & 0 \\ \hline
    \end{tabular}
    \label{tab:arch_sine_approx:resource_usage}
\end{table}

\subsection{Oscillator Virtualization} \label{sec:architecture:virtualization}

If the FPGA platform has a relative abundance of memory resources compared to computational resources, there is an opportunity to expand the network size by time-multiplexing the computation hardware using a memory banking\footnote{Also referred to as memory paging.} scheme.
With this scheme extra logic is added to allow each oscillator core to select a memory bank.
Each memory bank is a replica of the memory section seen in Fig. \ref{fig:architecture:oscillator_core}.
These replicas hold the coupling values of the virtual oscillators for that given core.
In this way, virtual oscillators are created by reusing the computation hardware over time but extending the memory hardware.
Naturally, each memory block has to be extended to allow the storage of the weights for the virtual oscillators.
Nevertheless, this memory-banking scheme allowed for a $4\times$ increase in network size, leading to a maximum total (virtual) oscillator count of $4\times128=512$, which is implented in ODEONN\footnote{We show the hardware utilization in Table \ref{tab:appendix:resource_usage} in the appendix.}.
The scheduling algorithm also needs to be modified slightly.
Algorithm \ref{alg:scheduler_banked} is an alternative version of the scheduling algorithm, which allows the use of memory banks in the oscillator cores.

\begin{algorithm}
    \caption{Algorithm for scheduling data transfer between cores with memory banks}
    \label{alg:scheduler_banked}
    \begin{algorithmic}
        \For {$N_\text{iterations}$}
            \For {$b = 1$ to $N_\text{banks}$}
                \For {$i = 1$ to $N_\text{cores}$}
                    \State Obtain $\phi_i$ from core $i$
                    \State Transmit $\phi_i$ to each core
                \EndFor
                \State Send signal to store to phase to bank $b$ and reset summation memory 
            \EndFor
            \State Send signal to update phase to each core
        \EndFor
        \State Raise flag to indicate end of computation
    \end{algorithmic}
\end{algorithm}

\section{Methods} \label{sec:methods}

For verification of ODEONN, four benchmarks will be utilized.
These benchmarks stem from the two primary target applications for ONNs; Combinatorial optimization and associative memory.
From the former category, max-cut and Sudoku puzzle solving, a subset of graph coloring, will be used, and from the latter, two associative memory tasks are tested, one with binary colors, the other with grayscale colors.
The goal of the benchmarks is to demonstrate how ODEONN compares relative to a floating-point GPU ONN simulation.
If the performance difference is small, it shows that ODEONN sufficiently approximates the theory and that it could be used to accelerate future benchmarks.
Hence, the achieved performance in this paper will not be compared to other works, since the relative performance between the theory and the architecture in this paper is the focus, not the performance of any given mapping or algorithm on ONN.
To this end, a Python-based software ODE Solver running on a GPU was utilized.
All simulation parameters, such as the time step size, total number of time steps, and the integration method were matched exactly to the parameters in the hardware architecture.
The only difference lies in the number representation, while ODEONN utilizes fixed-point quantization, the Python simulation fully utilizes 32-bit floating-point quantization.
This allows for a qualitative analysis of the effect of quantization for ODE solver-based ONN simulation, in hardware and software.
To analyze the effect of the sine approximation from Eq. \eqref{eq:arch_sine_approx:quadratic_approximation} in isolation from quantization effects, the Python simulations were run in several configurations per benchmark.
The configuration for each benchmark is shown in Table \ref{tab:methods:benchmarking_setup}.
\begin{table}
    \centering
    \begin{tabular}{|c|c|c|c|c|} \hline
        Benchmark & $\sin$ & $\sin + \widehat{W_{i,j}}$ & $\widehat{\sin}$ & $\widehat{\sin} + \widehat{W_{i,j}}$ \\ \hline
        Max-cut & $\checkmark$ & not applicable \tnote{a} & $\checkmark$ & not applicable \tnote{a} \\
        Sudoku & $\checkmark$ & not applicable \tnote{a}& $\checkmark$ & not applicable \tnote{a} \\
        Associative memory & $\checkmark$ & $\checkmark$ & $\checkmark$ & $\checkmark$ \\
        Grayscale Associative memory & $\checkmark$ & $\checkmark$ & $\checkmark$ & $\checkmark$ \\ \hline
    \end{tabular}
    \caption{Quantization and approximation setups used in Python per benchmark.}
    \label{tab:methods:benchmarking_setup}
    \begin{tablenotes}
        \item [a] The coupling values for these cases can be represented in the quantized form without loss of generality, thus these benchmarks would be redundant.
    \end{tablenotes}
\end{table}
Note that the effect of weight quantization is only analyzed for the associative memory benchmarks, since the weights can be represented exactly for the optimization problems even in the quantized form.

The Python simulations are run on a server running Red Hat Enterprise Linux version 8.10 utilizing an NVIDIA L40 GPU.
In addition to the benchmarks, the utilization of resources is checked by synthesizing ODEONN at $N_\text{cores}$ from $16$ up to and including $128$ and $N_\text{banks}$ from $1$ up to and including $8$.
The main takeaway of each benchmark will be highlighted in the results section.
The full results are provided in the appendix.

\subsection{Max-cut} \label{sec:methods:max_cut}

Max-cut is a combinatorial optimization problem, where a graph needs to be separated into two groups, while maximizing the number of edges between the groups.
The typical mapping of a max-cut problem to an ONN is to take the adjacency matrix of the graph as the weights between the oscillators, but set the weights to negative values, $-1$ for unweighted graphs, or minus the weight for weighted graphs \cite{lucas_ising_2014, wang_oim_2019}.
A data set was generated with node counts from $16$ to $512$, edge densities from $10\%$ to $75\%$.
For each node count and edge density $10$ random connected graphs were generated.
Each graph was evaluated $100$ times, each time with uniform random initial phases in the range $[0, 2\pi)$.
The average cut value was recorded at each data point for comparison.

\subsection{Sudoku Puzzles} \label{sec:methods:sudoku}

Sudoku is a subset of the graph coloring problem, where a grid of cells needs to be filled in with non-repeating digits in the sub-grids, columns, and rows.
The problem mapping introduced in \cite{Haverkort2026a} is used to benchmark ODEONN.
This mapping of Sudoku to ONN uses a non-Hermitian complex coupling matrix.
Thus, this gives a comparison point for the performance regarding problems that use complex coupling.
We repeat the methodology used in \cite{Haverkort2026a}, where solved puzzles were used to generate random input puzzles.
10 base puzzles are used, which are randomized 100 times, and simulated 10 times with different random initial phases.
To minimize the effect of run variance, we use the exact same input puzzles on ODEONN and on the GPU ONN.
Additionally, the GPU ONN is run with an exact sine function and with the approximation sine function to isolate the effect of the approximation from the results.

\subsection{Associative Memory} \label{sec:methods:associative_memory}

\begin{figure}
    \centering
    \begin{subfigure}{0.47\linewidth}
        \centering
        \includegraphics[width=\linewidth]{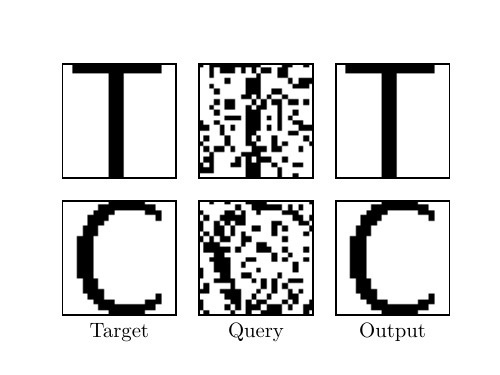}
        \caption{}
        \label{fig:methods:binary_example}
    \end{subfigure}
    \hfill
    \begin{subfigure}{0.52\linewidth}
        \centering
        \includegraphics[width=\linewidth]{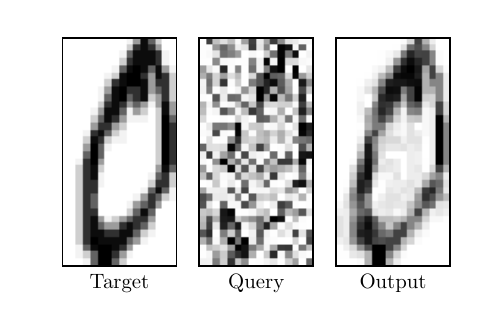}
        \caption{}
        \label{fig:methods:grayscale_example}
    \end{subfigure}
    \label{fig:methods:associative_memory}
    \caption{Pattern retrieval examples.\\(a) Example of pattern retrieval. On the left the target pattern is shown, which has been corrupted ($25\%$) to obtain the query in the middle. After performing pattern retrieval the output is obtained from ODEONN.\\(b) Example of grayscale pattern retrieval. On the left the target pattern is shown, which has noise added ($25\%$) to obtain the query in the middle. After performing pattern retrieval the output is obtained from ODEONN.}
\end{figure}

To evaluate performance in terms of associative memory tasks, a dataset of alphabet letters was trained using the Diederich-Opper 2 \cite{diederich_learning_1987} learning rule.
The weights are normalized to $[-127, 127]$, the full 8-bit range.
For each pattern learned in the dataset, $1000$ randomly corrupted versions were generated at three corruption levels; $10\%$, $20\%$, and $25\%$.
The corruption percentage is the number of pixels that get inverted, so either flipped from black to white, or from white to black.
To prevent dead dynamics at the initial conditions the equivalent of one LSB of noise was added, i.e. $\epsilon = 2^{-15}$.
These corrupted versions were used as queries for the associative memory tasks.
Then, for each query, the phase dynamics were simulated in the aforementioned configurations and the final phase values were evaluated after $5000$ integration steps.
A query is considered correct when the binarized phases, which were binarized using a simple boxing algorithm, were exactly equal to the expected target pattern, otherwise the pattern is deemed incorrect.
See Fig. \ref{fig:methods:binary_example} for an example of the three steps in an associative memory process.

\subsection{Grayscale Associative Memory} \label{sec:methods:associative_memory_grayscale}

To evaluate the functionality of the complex coupling functionality, a grayscale associative memory benchmark is used.
Similar to the black-and-white associative memory task, a dataset is constructed, instead this time two symbols are taken from the MNIST\cite{deng_mnist_2012} dataset.
The color space of the patterns is reduced such that there are 8 steps from fully white to fully black.
Following this, the complex coupling weights are computed using the complex Hebbian learning rule, as described in \cite{hoppensteadt_pattern_2000}.
The weight amplitudes are normalized to $[-127, 127]$, the full 8-bit range.
Additive uniform random noise will be used to disturb the images, instead of randomly corrupting, or flipping pixels as described in Section \ref{sec:methods:associative_memory}.
Again, for each pattern learned in the dataset, $1000$ versions were generated at three noise levels; $10\%$, $20\%$, and $25\%$.
The noise percentage is a percentage of the total amplitude range, $2\pi$.
For example, $10\%$ noise means that noise values are sampled from the range $[-0.2\pi, 0.2\pi]$. 
These noisy versions were used as queries for the associative memory tasks.
Then, for each query, the phase dynamics were simulated in the aforementioned configurations and the final phase values were evaluated after $5000$ integration steps.
Fig. \ref{fig:methods:grayscale_example} shows the three stages of a pattern in the grayscale associative memory process.
The accuracy of the results are evaluated using the structural similarity index measure (SSIM) \cite{Wang2004}, where a value of 1 means the patterns are identical and 0 means they are completely different.

\subsection{Performance and Efficiency} \label{sec:methods:speed_and_efficiency}

To obtain a performance metric for both platforms, a random weight matrix with weights in the range $[-1, 1]$ is used as a test.
By using a random weight matrix, any compiler optimization that sparsifies the matrix can be ruled out.
This weight matrix is then simulated for one million integration steps and the average integration steps per second is computed.
The average power draw of the L40 graphics processing unit (GPU) is taken from the NVIDIA System Management Interface.
No other processes were utilizing the GPU during testing to ensure all power consumption is caused by the ONN simulation.
The power consumption of the FPGA is taken as the power reported by the Vivado development software.

\section{Benchmarking Results} \label{sec:results}

\subsection{Max-cut} \label{sec:results:max_cut}

Table \ref{tab:results:max_cut} shows the results for the max-cut benchmark.
As can be seen for each edge density, the relative difference is below 0.1\%, which can be considered within run-to-run variance.
Therefore, neither the approximation of the sine function nor the fixed-point quantization of ODEONN has any effect on the performance of ONN for a max-cut task; on the contrary, the approximation of the sine function seems to have slightly improved the average cut values.
However, this improvement is almost negligible.

\begin{table}
    \captionsetup{width=\textwidth}
    \caption{Max-cut performance comparison for a 512 node graph.}
    \centering
    \begin{tabular}{|c|c|c|c|c|} \hline
        Implementation & \makecell{Average \\ Cut value [-]\\ $10\%$ density} & \makecell{Average \\ Cut value [-]\\ $25\%$ density} & \makecell{Average \\ Cut value [-]\\ $50\%$ density} & \makecell{Average \\ Cut value [-]\\ $75\%$ density} \\ \hline
        GPU ONN & 7719 (-0.1\%) & 18088 (-0.1\%) & 34729 (-0.1\%) & 50855 (-0.0\%) \\ 
        GPU ONN + $\widehat{\sin}$ & 7730 (0\%) & 18107 (+0.0\%) & 34754 (+0.0\%) & 50883 (+0.0\%)\\ 
        ODEONN & 7730 & 18105 & 34752 & 50880\\ \hline
    \end{tabular}
    \label{tab:results:max_cut}
    \textit{Note: The values in brackets are the relative difference to ODEONN.}
\end{table}

\subsection{Sudoku} \label{sec:results:sudoku}
\begin{figure}
    \centering
    \includegraphics[width=\linewidth]{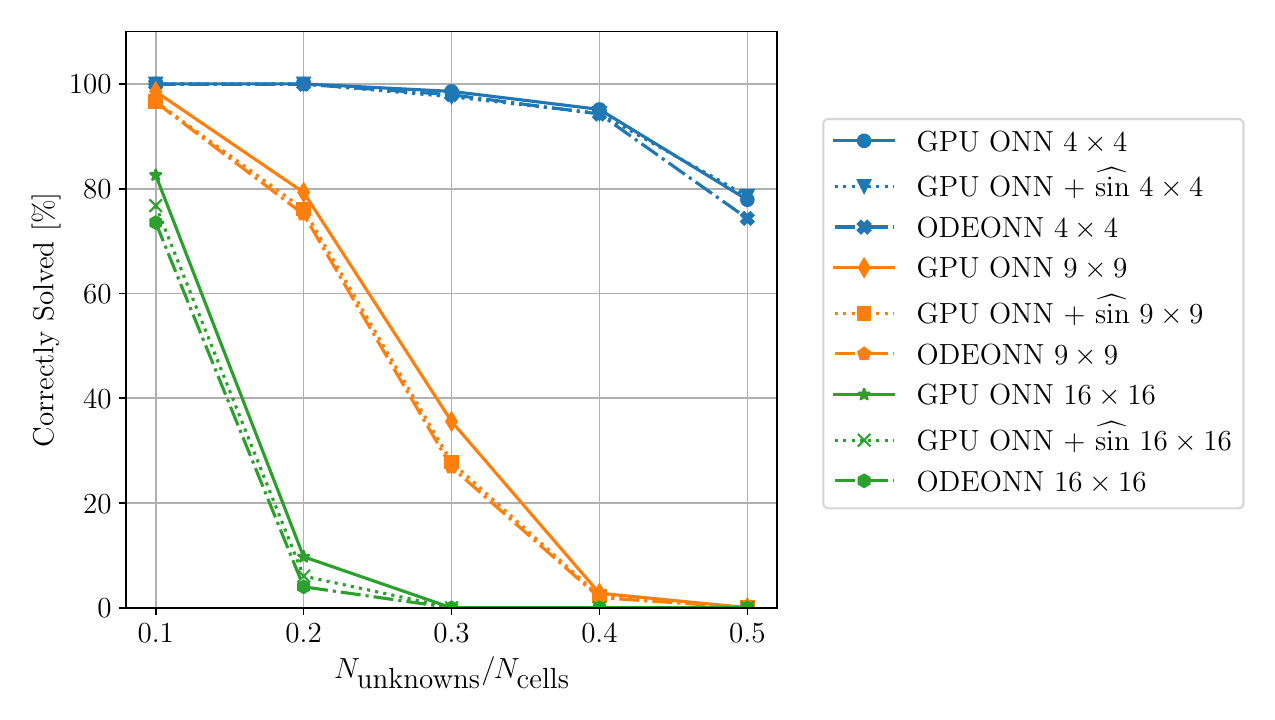}
    \caption{Sudoku retrieval rate across different puzzle sizes and ratios between the number of known and unknown cells.}
    \label{fig:results:sudoku}
\end{figure}

As seen in Figure \ref{fig:results:sudoku}, ODEONN performs close to the GPU ONN, with a slight reduction in performance of at most 9 percentage points.
This deviation can be explained entirely by the approximated sine, because it can be seen that the results of the software simulation with the approximated sine coincide closely with the results of ODEONN.
Therefore, it can be concluded that the quantization of the phases and weight values does not affect the accuracy of the benchmark, but the approximated waveform does.

\subsection{Associative Memory} \label{sec:results:associative_memory}

The results of the associative memory benchmark are shown in Table \ref{tab:results:associative_memory}.
As can be seen, ODEONN is outperformed by the GPU ONN when the GPU ONN implementation is using non-quantized weights.
However, when the GPU ONN also operates using quantized weights, the performance gap is closed.
Thus, the quantization of the trained weights has a large impact on associative memory performance.
Performance of the GPU implementation with and without the approximation of the sine function is nearly identical.
Therefore, the impact of approximating the sine wave is small for the associative memory task.
Comparing the GPU ONN results with and without weight quantization reveals that most of the performance reduction can be attributed to the quantization of the weights.
This is in line with observations of previous work \cite{abernot_oscillatory_2023}.
Therefore, it can be concluded that the quantization of the phase quantities is sufficient to perform associative memory tasks, but that those of the coupling values are insufficient.
If one is specifically targeting this application or similar ones, it is recommended to increase the number of bits for the weight amplitude $W_{i,j}$.

\begin{table}
    \captionsetup{width=\textwidth}
    \caption{Associative memory performance comparison for $22\times22$ patterns.}
    \centering
    \begin{tabular}{|c|c|c|c|} \hline
        Implementation & \makecell{Correct [\%]\\ $10\%$ Corrupted} & \makecell{Correct [\%]\\ $20\%$ Corrupted} & \makecell{Correct [\%]\\ $25\%$ Corrupted} \\ \hline
        GPU ONN & 100 (+25\%) & 99.70 (+45\%) & 90.02 (+47\%) \\ 
        \makecell{GPU ONN + $\widehat{W_{i,j}}$} & 66.18 (-17\%) & 61.88 (-9.7\%) & 57.86 (-5.3\%) \\
        GPU ONN + $\widehat{\sin}$ & 100 (+25\%) & 100 (+46\%) & 88.60 (+45\%)\\ 
        \makecell{GPU ONN + $\widehat{\sin}$ +\\$\widehat{W_{i,j}}$} & 74.72 (-6.4\%) & 63.58 (-7.2\%) & 59.30 (-2.9\%)\\ 
        ODEONN & 79.78 & 68.54 & 61.08 \\ \hline
    \end{tabular}
    \label{tab:results:associative_memory}
    \textit{Note: The values in brackets are the relative difference to ODEONN.}
\end{table}

\begin{figure}
    \centering
    \includegraphics[width=0.5\linewidth]{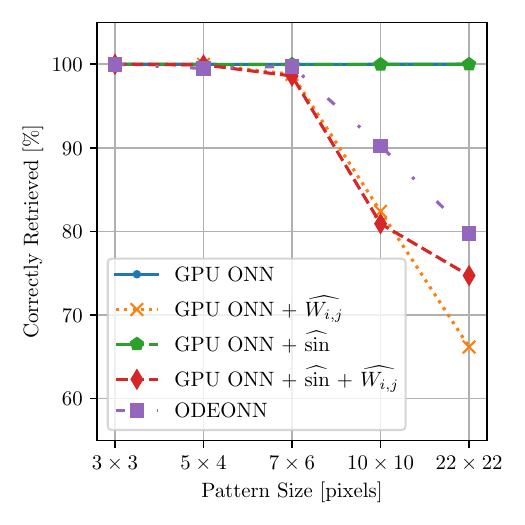}
    \caption{Pattern retrieval accuracy for various pattern sizes with 10\% pixel corruption.}
    \label{fig:results:associative_memory}
\end{figure}

\subsection{Grayscale Associative Memory} \label{sec:results:associative_memory_grayscale}

As seen in Table \ref{tab:results:associative_memory_grayscale}, the quantization of the weight values does not seem to have any effect on the outcome of the computation, in contrast to the results for binary associative memory.
This is likely due to the fact that patterns are already limited to 8 colors and only two patterns are trained\footnote{Only two patterns are trained because the learning rule was not capable of memorizing more without breaking down. This is a limitation encountered in the complex Hebbian learning rule, which is outside the scope of this paper.}, which typically requires less granularity in the weight values.
There is, however, a measurable difference when the approximation of the sine function is used.
In this case, the SSIM is lower than with an exact sine function.
A possible explanation for this is that the error shown in Fig. \ref{fig:arch_sine_approx:error} slightly displaces the relative stable points between two grayscale values, therefore causes a slight color difference in pixels, which affects the SSIM.
The performance difference is only $2\%$, thus the ODEONN architecture still performs very close to the GPU ONN simulation.

\begin{table}
    \captionsetup{width=\textwidth}
    \caption{Grayscale associative memory performance comparison.}
    \centering
    \begin{tabular}{|c|c|c|c|} \hline
        Implementation & \makecell{Mean SSIM [-]\\ $10\%$ Noise} & \makecell{Mean SSIM [-]\\ $20\%$ Noise} & \makecell{Mean SSIM [-]\\ $25\%$ Noise} \\ \hline
        GPU ONN & 0.92950 (+2.0\%) & 0.92950 (+2.0\%) & 0.92950 (+2.0\%) \\ 
        \makecell{GPU ONN + $\widehat{W_{i,j}}$} & 0.92954 (+2.0\%) & 0.92954 (+2.0\%) & 0.92954 (+2.0\%)\\
        GPU ONN + $\widehat{\sin}$ & 0.91130 (-0.0\%) & 0.91130 (-0.0\%) & 0.91130 (-0.0\%)\\ 
        \makecell{GPU ONN + $\widehat{\sin}$ +\\$\widehat{W_{i,j}}$} & 0.91136 (-0.0\%) & 0.91136 (-0.0\%) & 0.91136 (-0.0\%)\\ 
        ODEONN & 0.91138 & 0.91138 & 0.91138\\ \hline
    \end{tabular}
    \label{tab:results:associative_memory_grayscale}
    \textit{Note: The values in brackets are the relative difference to ODEONN.}
\end{table}

\subsection{Performance and Efficiency} \label{sec:results:runtime}

As can be seen in Table \ref{tab:results:runtime}, ODEONN architecture computes about $15\%$ more iterations per second than the software implementation running on GPU.
During the simulation of the ONN, the FPGA consumes $0.029\times$ the power of the GPU.
This means that not only is ODEONN faster, it also consumes less power, leading to a total improvement of the Energy-Delay Product of over $45\times$.
Of course, the GPU would be capable of simulating larger networks, while ODEONN is limited to a network of 512 on this particular board.

\begin{table}
    \centering
    \captionsetup{width=\textwidth}
    \caption{Performance comparison ODEONN versus GPU for a 512 oscillator network}
    \label{tab:results:runtime}
    \begin{tabular}{|c|c|c|c|}
        \hline
        Implementation      & Power [W]         & Throughput [it/s]          & EDP [nJs]         \\ \hline
        ODEONN              & 2.8               & 48285                 & 1.2               \\
        GPU ONN             & 95                & 41878                 & 54                \\ \hline
        Ratio (ODEONN/GPU)  & 0.029             & 1.15                  & 0.022             \\ \hline
    \end{tabular}
    \textit{Note: The Energy-Delay Product is computed as $\text{EDP} = P\cdot(\frac{1}{R})^2$, where $P$ is the power and $R$ is the throughput.}
\end{table}

\subsection{Resource Utilization} \label{sec:results:resource_usage}

For reference, the resource utilization without memory banking is shown in Table \ref{tab:results:resource_usage} and the resource usage of the final design used for benchmarking is shown in Table \ref{tab:results:resource_usage_complete}.
Roughly linear scaling of each resource can be observed up to the point where the maximum number of DSP cores is exceeded.
We found that the maximum network size on the PYNQ-Z2 that can be achieved is 512 oscillators.
The exact configurations of $N_\text{banks}$ and $N_\text{cores}$ can be found in the appendix.

\begin{table}
    \centering
    \captionsetup{width=\textwidth}
    \caption{Resource usage without memory banking excluding AXI interfaces.}
    \label{tab:results:resource_usage}
    \begin{tabular}{|c|c|c|c|c|}
        \hline
        $N_\text{cores}$ & 16 & 32 & 64 & 128\tnote{a} \\ \hline
        LUTs    & 2146  & 4280  & 8732  & 25606 \\
        FFs     & 1875  & 3672  & 7278  & 17649 \\
        DSPs    & 32    & 64    & 128   & 220 \\
        BRAMs   & 8     & 16    & 32    & 64 \\ \hline
    \end{tabular}
    \begin{tablenotes}
        \item [a] At this point, the maximum number of DSPs provided by the FPGA is exceeded, the remaining logic has been absorbed into LUTs and FFs.
    \end{tablenotes}
\end{table}

\begin{table}
    \centering
    \captionsetup{width=\textwidth}
    \caption{Resource utilization of final design excluding AXI interfaces $N_\text{cores} = 128$, $N_\text{banks}=4$ leading to $512$ virtual oscillators.}
    \begin{tabular}{|c|c|c|c|}
        \hline
        LUTs & FFs & BRAMs & DSPs \\ \hline
        36853 & 29960 & 128 & 220\tnote{a} \\ \hline
    \end{tabular}
    \label{tab:results:resource_usage_complete}
    \begin{tablenotes}
        \item [a] This is the maximum number of DSPs provided by the FPGA, the remaining logic has been absorbed into LUTs and FFs.
    \end{tablenotes}
\end{table}

\section{Conclusion} \label{sec:conclusion}

In this paper an FPGA-based ODE Solver architecture for Oscillatory Neural Networks called ODEONN has been described.
This architecture has been benchmarked using four typical applications tackled using ONNs, two NP optimization tasks, and two associative memory tasks.
For each of these tasks, either no performance penalty or a penalty of less than $2\%$ has been observed, when compared to a full precision software simulation running on GPU.
Depending on the benchmark, the performance penalty came from the sine approximation or from the weight quantization, the latter of which can easily be resolved by increasing the number bits.
In some cases, ODEONN even outperformed the simulation.
Furthermore, this architecture showed a $45\times$ energy-delay product increase when compared to a GPU-based ONN simulation.

\section{Discussion} \label{sec:discussion}

One main limitation of ODEONN is the communication overhead of transmitting the phase data between each core will increase linearly with the amount of cores, since the phase data is transmitted serially in time, but in parallel to all cores.
Thus, as the architecture is scaled up, the time spent transmitting will also go up.
However, the ratio between transmission and computing will remain the same, regardless of the number of cores.
For each of the $N_\text{cores}$, there is the same number of $\phi_i$ that has to be transmitted to the other cores and each additional $\phi_i$ takes an extra clock cycle to process.
There is no immediate solution that would resolve this communication overhead, without significantly increasing the usage of computing resources.
For example, one could consider transmitting multiple phases in parallel to reduce the overhead, but then multiple phases would also need to be processed in parallel, or buffered if computed in serial.

\section{Data Availability}

The data that support the findings of this study are available from the corresponding author upon reasonable request.

\section{Acknowledgment}

This work has received funding from the Dutch Research Council`s AiNed Fellowship research program, AI-on-ONN project under grant agreement No. NGF.1607.22.016.

\section {Author Contributions}

Bram F. Haverkort: Conceptualization, Data curation, Formal analysis, Investigation, Methodology, Software, Validation, Visualization, Writing - Original Draft;
Aida Todri-Sanial: Conceptualization, Funding acquisition, Project administration, Resources, Supervision, Writing - Review \& Editing; 

\section{Competing interests}

Bram F. Haverkort and Aida Todri-Sanial are inventors on patent application NL4002111 filed by the Eindhoven University of Technology, covering the contents of this paper.

\bibliography{library_b_haverkort}

@article{abernot_digital_2021,
	title = {Digital {Implementation} of {Oscillatory} {Neural} {Network} for {Image} {Recognition} {Applications}},
	volume = {15},
	issn = {1662-453X},
	url = {https://www.frontiersin.org/journals/neuroscience/articles/10.3389/fnins.2021.713054/full},
	doi = {10.3389/fnins.2021.713054},
	language = {English},
	urldate = {2024-06-21},
	journal = {Frontiers in Neuroscience},
	author = {Abernot, Madeleine and Gil, Thierry and Jiménez, Manuel and Núñez, Juan and Avellido, María J. and Linares-Barranco, Bernabé and Gonos, Théophile and Hardelin, Tanguy and Todri-Sanial, Aida},
	month = aug,
	year = {2021},
	note = {Publisher: Frontiers},
}

@article{wang_solving_2021,
	title = {Solving combinatorial optimisation problems using oscillator based {Ising} machines},
	volume = {20},
	issn = {1572-9796},
	url = {https://doi.org/10.1007/s11047-021-09845-3},
	doi = {10.1007/s11047-021-09845-3},
	language = {en},
	number = {2},
	urldate = {2024-06-21},
	journal = {Natural Computing},
	author = {Wang, Tianshi and Wu, Leon and Nobel, Parth and Roychowdhury, Jaijeet},
	month = jun,
	year = {2021},
	pages = {287--306},
}

@inproceedings{landge_n-oscillator_2020,
	title = {n-{Oscillator} {Neural} {Network} based {Efficient} {Cost} {Function} for n-city {Traveling} {Salesman} {Problem}},
	url = {https://ieeexplore.ieee.org/abstract/document/9206856},
	doi = {10.1109/IJCNN48605.2020.9206856},
	urldate = {2024-06-21},
	booktitle = {2020 {International} {Joint} {Conference} on {Neural} {Networks} ({IJCNN})},
	author = {Landge, Shruti and Saraswat, Vivek and Singh, Srisht Fateh and Ganguly, Udayan},
	month = jul,
	year = {2020},
	note = {ISSN: 2161-4407},
	pages = {1--8},
}

@article{vadlamani_combinatorial_2024,
	title = {Combinatorial optimization using the {Lagrange} primal-dual dynamics of parametric oscillator networks},
	volume = {21},
	doi = {10.1103/PhysRevApplied.21.044042},
	number = {4},
	journal = {Physical Review Applied},
	author = {Vadlamani, Sri Krishna},
	year = {2024},
}

@inproceedings{wang_oim_2019,
	address = {Cham},
	title = {{OIM}: {Oscillator}-{Based} {Ising} {Machines} for {Solving} {Combinatorial} {Optimisation} {Problems}},
	isbn = {978-3-030-19311-9},
	shorttitle = {{OIM}},
	doi = {10.1007/978-3-030-19311-9_19},
	language = {en},
	booktitle = {Unconventional {Computation} and {Natural} {Computation}},
	publisher = {Springer International Publishing},
	author = {Wang, Tianshi and Roychowdhury, Jaijeet},
	editor = {McQuillan, Ian and Seki, Shinnosuke},
	year = {2019},
	pages = {232--256},
}

@article{hoppensteadt_pattern_2000,
	title = {Pattern recognition via synchronization in phase-locked loop neural networks},
	volume = {11},
	issn = {1941-0093},
	url = {https://ieeexplore.ieee.org/document/846744},
	doi = {10.1109/72.846744},
	number = {3},
	urldate = {2024-06-24},
	journal = {IEEE Transactions on Neural Networks},
	author = {Hoppensteadt, F.C. and Izhikevich, E.M.},
	month = may,
	year = {2000},
	note = {Conference Name: IEEE Transactions on Neural Networks},
	pages = {734--738},
}

@article{abernot_oscillatory_2023,
	title = {Oscillatory neural network learning for pattern recognition: an on-chip learning perspective and implementation},
	volume = {17},
	issn = {1662-453X},
	shorttitle = {Oscillatory neural network learning for pattern recognition},
	url = {https://www.frontiersin.org/journals/neuroscience/articles/10.3389/fnins.2023.1196796/full},
	doi = {10.3389/fnins.2023.1196796},
	language = {English},
	urldate = {2024-07-04},
	journal = {Frontiers in Neuroscience},
	author = {Abernot, Madeleine and Azemard, Nadine and Todri-Sanial, Aida},
	month = jun,
	year = {2023},
	note = {Publisher: Frontiers},
}

@article{lucas_ising_2014,
	title = {Ising formulations of many {NP} problems},
	volume = {2},
	issn = {2296-424X},
	url = {http://arxiv.org/abs/1302.5843},
	doi = {10.3389/fphy.2014.00005},
	urldate = {2024-07-04},
	journal = {Frontiers in Physics},
	author = {Lucas, Andrew},
	year = {2014},
	note = {arXiv:1302.5843 [cond-mat, physics:quant-ph]},
}

@inproceedings{sabo_classonn_2024,
	address = {Valencia, Spain},
	title = {{ClassONN}: {Classification} with {Oscillatory} {Neural} {Networks} {Using} the {Kuramoto} {Model}},
	shorttitle = {{ClassONN}},
	url = {https://ieeexplore.ieee.org/abstract/document/10546829},
	urldate = {2024-07-04},
	booktitle = {2024 {Design}, {Automation} \& {Test} in {Europe} {Conference} \& {Exhibition} ({DATE})},
	publisher = {IEEE},
	author = {Sabo, Filip and Todri-Sanial, Aida},
	month = mar,
	year = {2024},
	note = {ISSN: 1558-1101},
	pages = {1--2},
}

@inproceedings{bashar_fpga-based_2024,
	address = {San Francisco, CA, USA},
	title = {An {FPGA}-based {Max}-{K}-{Cut} {Accelerator} {Exploiting} {Oscillator} {Synchronization} {Model}},
	url = {https://ieeexplore.ieee.org/document/10528742},
	doi = {10.1109/ISQED60706.2024.10528742},
	urldate = {2024-07-19},
	booktitle = {2024 25th {International} {Symposium} on {Quality} {Electronic} {Design} ({ISQED})},
	publisher = {IEEE},
	author = {Bashar, Mohammad Khairul and Li, Zheyu and Narayanan, Vijaykrishnan and Shukla, Nikhil},
	month = apr,
	year = {2024},
	note = {ISSN: 1948-3295},
	pages = {1--8},
}

@article{diederich_learning_1987,
	title = {Learning of correlated patterns in spin-glass networks by local learning rules},
	volume = {58},
	url = {https://link.aps.org/doi/10.1103/PhysRevLett.58.949},
	doi = {10.1103/PhysRevLett.58.949},
	number = {9},
	urldate = {2024-09-16},
	journal = {Physical Review Letters},
	author = {Diederich, Sigurd and Opper, Manfred},
	month = mar,
	year = {1987},
	note = {Publisher: American Physical Society},
	pages = {949--952},
}

@article{deng_mnist_2012,
	title = {The {MNIST} {Database} of {Handwritten} {Digit} {Images} for {Machine} {Learning} {Research} [{Best} of the {Web}]},
	volume = {29},
	issn = {1558-0792},
	url = {https://ieeexplore.ieee.org/document/6296535},
	doi = {10.1109/MSP.2012.2211477},
	number = {6},
	urldate = {2024-09-16},
	journal = {IEEE Signal Processing Magazine},
	author = {Deng, Li},
	month = nov,
	year = {2012},
	note = {Conference Name: IEEE Signal Processing Magazine},
	pages = {141--142},
}

@inproceedings{sreedhara_digital_2023,
	title = {Digital {Emulation} of {Oscillator} {Ising} {Machines}},
	url = {https://ieeexplore.ieee.org/document/10137084},
	doi = {10.23919/DATE56975.2023.10137084},
	urldate = {2024-09-25},
	booktitle = {2023 {Design}, {Automation} \& {Test} in {Europe} {Conference} \& {Exhibition} ({DATE})},
	author = {Sreedhara, Shreesha and Roychowdhury, Jaijeet and Wabnig, Joachim and Srinath, Pavan},
	month = apr,
	year = {2023},
	note = {ISSN: 1558-1101},
	pages = {1--2},
}

@inproceedings{delacour_oscillatory_2021,
	address = {Tampa, FL, USA},
	title = {Oscillatory {Neural} {Networks} for {Edge} {AI} {Computing}},
	url = {https://ieeexplore.ieee.org/document/9516738},
	doi = {10.1109/ISVLSI51109.2021.00066},
	urldate = {2024-11-08},
	booktitle = {2021 {IEEE} {Computer} {Society} {Annual} {Symposium} on {VLSI} ({ISVLSI})},
	publisher = {IEEE},
	author = {Delacour, Corentin and Carapezzi, Stefania and Abernot, Madeleine and Boschetto, Gabriele and Azemard, Nadine and Salles, Jeremie and Gil, Thierry and Todri-Sanial, Aida},
	month = jul,
	year = {2021},
	note = {ISSN: 2159-3477},
	pages = {326--331},
}

@article{todri-sanial_computing_2024,
	title = {Computing with oscillators from theoretical underpinnings to applications and demonstrators},
	volume = {1},
	copyright = {2024 The Author(s)},
	issn = {3004-8672},
	url = {https://www.nature.com/articles/s44335-024-00015-z},
	doi = {10.1038/s44335-024-00015-z},
	language = {en},
	number = {1},
	urldate = {2024-12-04},
	journal = {npj Unconventional Computing},
	author = {Todri-Sanial, Aida and Delacour, Corentin and Abernot, Madeleine and Sabo, Filip},
	month = dec,
	year = {2024},
	note = {Publisher: Nature Publishing Group},
	pages = {1--16},
}

@inproceedings{su_roc-spin_2024,
	title = {{ROC}-{Spin}: {A} 28nm 2,000 {Ring}-{Oscillator}-{Collapse} {Spins} for {Solving} {Combinatorial} {Optimization} {Problems}},
	shorttitle = {{ROC}-{Spin}},
	url = {https://ieeexplore.ieee.org/abstract/document/10848963/authors},
	doi = {10.1109/A-SSCC60305.2024.10848963},
	urldate = {2025-02-03},
	booktitle = {2024 {IEEE} {Asian} {Solid}-{State} {Circuits} {Conference} ({A}-{SSCC})},
	author = {Su, Yuqi and Do, Anh Tuan and Kim, Tony Tae-Hyoung and Kim, Bongjin},
	month = nov,
	year = {2024},
	pages = {1--3},
}

@article{li_improved_2025,
	title = {Improved {Mapping} {Strategies} for {Integrated} {Oscillator}-{Based} {Ising} {Machine} {Networks}},
	volume = {95},
	doi = {10.54254/2753-8818/2024.21348},
	journal = {Theoretical and Natural Science},
	author = {Li, Boyang},
	month = mar,
	year = {2025},
	pages = {152--159},
}

@misc{gower_how_2025,
	title = {How to {Train} an {Oscillator} {Ising} {Machine} using {Equilibrium} {Propagation}},
	url = {http://arxiv.org/abs/2505.02103},
	doi = {10.48550/arXiv.2505.02103},
	urldate = {2025-05-08},
	publisher = {arXiv},
	author = {Gower, Alex},
	month = may,
	year = {2025},
	note = {arXiv:2505.02103 [cond-mat]},
}

@article{soni_spinonn_2025,
	title = {{SpinONN}: energy efficient brain-inspired spintronics-based {Hopfield} oscillatory neural network for image denoising},
	volume = {5},
	issn = {2634-4386},
	shorttitle = {{SpinONN}},
	url = {https://dx.doi.org/10.1088/2634-4386/ade622},
	doi = {10.1088/2634-4386/ade622},
	language = {en},
	number = {3},
	urldate = {2025-06-30},
	journal = {Neuromorphic Computing and Engineering},
	author = {Soni, Sandeep and Rezaeiyan, Yasser and Boehnert, Tim and Farkhani, Hooman and Ferreira, Ricardo and Kaushik, Brajesh Kumar and Moradi, Farshad and Shreya, Sonal},
	month = jun,
	year = {2025},
	note = {Publisher: IOP Publishing},
	pages = {034001},
}

@misc{cai_oscnet_2025,
	title = {{OscNet} v1.5: {Energy} {Efficient} {Hopfield} {Network} on {CMOS} {Oscillators} for {Image} {Classification}},
	shorttitle = {{OscNet} v1.5},
	url = {http://arxiv.org/abs/2506.12610},
	doi = {10.48550/arXiv.2506.12610},
	urldate = {2025-06-30},
	publisher = {arXiv},
	author = {Cai, Wenxiao and Li, Zongru and Wang, Iris and Wang, Yu-Neng and Lee, Thomas H.},
	month = jun,
	year = {2025},
	note = {arXiv:2506.12610 [cs]},
}

@article{cilasun_coupled-oscillator-based_2025,
	title = {A coupled-oscillator-based {Ising} chip for combinatorial optimization},
	volume = {8},
	copyright = {2025 The Author(s), under exclusive licence to Springer Nature Limited},
	issn = {2520-1131},
	url = {https://www.nature.com/articles/s41928-025-01393-3},
	doi = {10.1038/s41928-025-01393-3},
	language = {en},
	number = {6},
	urldate = {2025-06-30},
	journal = {Nature Electronics},
	author = {Cılasun, Hüsrev and Moy, William and Zeng, Ziqing and Islam, Tahmida and Lo, Hao and Vanasse, Alex and Tan, Megan and Anees, Mohammad and S, Ramprasath and Kumar, Abhimanyu and Sapatnekar, Sachin S. and Kim, Chris H. and Karpuzcu, Ulya R.},
	month = jun,
	year = {2025},
	note = {Publisher: Nature Publishing Group},
	pages = {537--546},
}

@misc{gonul_gpu-accelerated_2025,
	title = {{GPU}-{Accelerated} {Simulated} {Oscillator} {Ising}/{Potts} {Machine} {Solving} {Combinatorial} {Optimization} {Problems}},
	url = {http://arxiv.org/abs/2505.22631},
	doi = {10.48550/arXiv.2505.22631},
	urldate = {2025-06-30},
	publisher = {arXiv},
	author = {Gonul, Yilmaz Ege and Kayan, Ceyhun Efe and Mustafazade, Ilknur and Kandasamy, Nagarajan and Taskin, Baris},
	month = may,
	year = {2025},
	note = {arXiv:2505.22631 [cs]},
}

@inproceedings{liu_efficient_2024,
	title = {An {Efficient} {Simulated} {Oscillator}-{Based} {Ising} {Machine} on {FPGAs}},
	url = {https://ieeexplore.ieee.org/document/10628658},
	doi = {10.1109/NANO61778.2024.10628658},
	urldate = {2025-08-22},
	booktitle = {2024 {IEEE} 24th {International} {Conference} on {Nanotechnology} ({NANO})},
	author = {Liu, Bailiang and Zhang, Tingting and Gao, Xingjian and Han, Jie},
	month = jul,
	year = {2024},
	note = {ISSN: 1944-9380},
	pages = {469--474},
}

@book{Kuramoto_1984, address={Berlin, Heidelberg}, series={Springer Series in Synergetics}, title={Chemical Oscillations, Waves, and Turbulence}, volume={19}, rights={http://www.springer.com/tdm}, ISBN={978-3-642-69691-6}, url={http://link.springer.com/10.1007/978-3-642-69689-3}, DOI={10.1007/978-3-642-69689-3}, publisher={Springer}, author={Kuramoto, Yoshiki}, year={1984}, collection={Springer Series in Synergetics} }

@article{Haverkort2026a,
  title = {Solving {{Sudoku}} Using Oscillatory Neural Networks},
  author = {Haverkort, Bram F and Sbravati, Federico and Porfir, Stefan and {Todri-Sanial}, Aida},
  year = 2026,
  month = may,
  journal = {Neuromorphic Computing and Engineering},
  volume = {6},
  number = {2},
  pages = {024007},
  publisher = {IOP Publishing},
  issn = {2634-4386},
  doi = {10.1088/2634-4386/ae65d4},
  urldate = {2026-07-09},
  langid = {english}
}

@misc{Gonul2026,
  title = {An {{ASIC Emulated Oscillator Ising}}/{{Potts Machine Solving Combinatorial Optimization Problems}}},
  author = {Gonul, Yilmaz Ege and Taskin, Baris},
  year = 2026,
  month = apr,
  number = {arXiv:2604.14027},
  eprint = {2604.14027},
  primaryclass = {cs},
  publisher = {arXiv},
  doi = {10.48550/arXiv.2604.14027},
  urldate = {2026-04-20},
  archiveprefix = {arXiv}
}

@inproceedings{DeepGoru2026,
  title = {{{DOIM140}}: {{A}} 140-Spin, All-to-All Connected, {{Oscillator Ising Machine IC}} in 180nm {{CMOS}}},
  shorttitle = {{{DOIM140}}},
  booktitle = {2026 IEEE Custom Integrated Circuits Conference (CICC)},
  author = {Deep Goru, A. Gagan and Shreesha Sreedhara, B. and Pavan Sumanth Sikhakollu, C. Venkata and Thomas Jagielski, D. and Manohar, E. Rajit and Jaijeet Roychowdhury, F.},
  year = 2026,
  month = apr,
  pages = {1--4},
  issn = {2152-3630},
  doi = {10.1109/CICC65509.2026.11509608},
  urldate = {2026-06-30}
}

@article{Wang2004,
  title = {Image Quality Assessment: From Error Visibility to Structural Similarity},
  shorttitle = {Image Quality Assessment},
  author = {Wang, Zhou and Bovik, A.C. and Sheikh, H.R. and Simoncelli, E.P.},
  year = 2004,
  month = apr,
  journal = {IEEE Transactions on Image Processing},
  volume = {13},
  number = {4},
  pages = {600--612},
  issn = {1941-0042},
  doi = {10.1109/TIP.2003.819861},
  urldate = {2026-07-28}
}

@inproceedings{Volder1959,
author = {Volder, Jack},
title = {The CORDIC computing technique},
year = {1959},
isbn = {9781450378659},
publisher = {Association for Computing Machinery},
address = {New York, NY, USA},
url = {https://doi.org/10.1145/1457838.1457886},
doi = {10.1145/1457838.1457886},
booktitle = {Papers Presented at the the March 3-5, 1959, Western Joint Computer Conference},
pages = {257–261},
numpages = {5},
location = {San Francisco, California},
series = {IRE-AIEE-ACM '59 (Western)}
}

@book{Euler1768,
address = {St Petersburg},
author = {Euler, Leonhard},
title = {Institutiones calculi integralis (Foundations of integral calculus)},
publisher = {Imperial Academy of Sciences},
chapter = {7},    
url = {https://archive.org/details/institutionescal020326mbp},
year = {1768},
number={1},
series={Institutionum calculi integralis},
}

\begin{appendices}

\section{Additional benchmarking data}\label{sec:appendix:extra_data}

The full benchmarking results are available in Tables \ref{tab:appendix:max_cut} and \ref{tab:appendix:associative_memory}.
ODEONN resource usage for all values of $N_\text{cores}$ and $N_\text{banks}$ is given in Table \ref{tab:appendix:resource_usage}.

\begin{table}
    \captionsetup{width=\textwidth}
    \caption{Max-cut performance comparison (Full table).}
    \centering
    \begin{tabular}{|c|c|c|c|c|} \hline
        Implementation & \makecell{Average \\ Cut value [-]\\ $10\%$ density} & \makecell{Average \\ Cut value [-]\\ $25\%$ density} & \makecell{Average \\ Cut value [-]\\ $50\%$ density} & \makecell{Average \\ Cut value [-]\\ $75\%$ density} \\ \hline
        \multicolumn{5}{|c|}{16 Nodes} \\ \hline
        GPU ONN & 16.9 (-0.6\%) & 24.4 (-0.8\%) & 39.7 (0\%) & 56.1 (-0.7\%) \\ 
        GPU ONN + $\widehat{\sin}$ & 16.9 (-0.6\%) & 24.4 (-0.8\%) & 39.7 (0\%) & 56.1 (-0.7\%) \\ 
        ODEONN & 17.0 & 24.6 & 39.7 & 56.5 \\ \hline
        \multicolumn{5}{|c|}{32 Nodes} \\ \hline
        GPU ONN & 45.0 (-0.4\%) & 85.3 (-0.6\%) & 157.8 (-0.3\%)& 214.8 (-0.2\%) \\ 
        GPU ONN + $\widehat{\sin}$ & 45.1 (-0.2\%) & 85.3 (-0.6\%) & 157.7 (-0.4\%) & 214.7 (-0.2\%) \\ 
        ODEONN & 45.2 & 85.8 & 158.3 & 215.2\\ \hline
        \multicolumn{5}{|c|}{64 Nodes} \\ \hline
        GPU ONN & 152.8 (-0.3\%) & 332.4 (-0.2\%) & 594.7 (-0.2\%) & 838.8 (-0.2\%) \\ 
        GPU ONN + $\widehat{\sin}$ & 152.9 (-0.3\%) & 332.3 (-0.2\%) & 594.4 (-0.3\%) & 838.7 (-0.2\%)\\ 
        ODEONN & 153.3 & 332.9 & 596.0 & 840.3 \\ \hline
        \multicolumn{5}{|c|}{128 Nodes} \\ \hline
        GPU ONN & 538.7 (-0.3\%) & 1235 (0.2\%) & 2277 (-0.2\%) & 3279 (-0.1\%)\\ 
        GPU ONN + $\widehat{\sin}$ & 540.3 (0\%) & 1238 (0\%) & 2281 (0\%) & 3282 (0\%) \\ 
        ODEONN & 540.3 & 1238 & 2281 & 3282 \\ \hline
        \multicolumn{5}{|c|}{256 Nodes} \\ \hline
        GPU ONN & 2048 (-0.2\%) & 4703 (-0.2\%) & 8872 (-0.1\%) & 12872 (-0.1\%) \\ 
        GPU ONN + $\widehat{\sin}$ & 2052 (0\%) & 4710 (-0.0\%) & 8882 (0\%) & 12884 (+0.0\%)\\ 
        ODEONN & 2052 & 4711 & 8882 & 12883\\ \hline
        \multicolumn{5}{|c|}{512 Nodes} \\ \hline
        GPU ONN & 7719 (-0.1\%) & 18088 (-0.1\%) & 34729 (-0.1\%) & 50855 (-0.0\%) \\ 
        GPU ONN + $\widehat{\sin}$ & 7730 (0\%) & 18107 (+0.0\%) & 34754 (+0.0\%) & 50883 (+0.0\%)\\ 
        ODEONN & 7730 & 18105 & 34752 & 50880\\ \hline
    \end{tabular}
    \label{tab:appendix:max_cut}
    \begin{tablenotes}
        \item \textit{Note: The values in brackets are the relative difference to ODEONN.}
    \end{tablenotes}
\end{table}

\begin{table}
    \captionsetup{width=\textwidth}
    \caption{Associative memory performance comparison (Full table).}
    \centering
    \begin{tabular}{|c|c|c|c|} \hline
        Implementation & \makecell{Correct [\%]\\ $10\%$ Corrupted} & \makecell{Correct [\%]\\ $20\%$ Corrupted} & \makecell{Correct [\%]\\ $25\%$ Corrupted} \\ \hline
        \multicolumn{4}{|c|}{$3\times3$ Patterns} \\ \hline
        GPU ONN & 100 (0\%) & 80.3 (+12\%)& 78.95 (+10\%) \\
        \makecell{GPU ONN +$\widehat{W_{i,j}}$} & 100 (0\%)& 81.40 (+13\%) & 79.25 (+11\%)\\
        GPU ONN + $\widehat{\sin}$ & 100 (0\%)& 80.45 (+12\%) & 79.15 (+11\%)\\ 
        \makecell{GPU ONN + $\widehat{\sin}$ +$\widehat{W_{i,j}}$} & 100 (0\%)& 80.90 (+13\%) & 79.15 (+11\%)\\ 
        ODEONN & 100 & 71.80 & 71.50\\ \hline
        \multicolumn{4}{|c|}{$5\times4$ Patterns} \\ \hline
        GPU ONN & 100 (+0.5\%) & 89.85 (+8.1\%) & 69.85 (+12\%)\\ 
        \makecell{GPU ONN +$\widehat{W_{i,j}}$} & 100 (+0.5\%) & 87.15 (+4.9\%) & 66.70 (+6.8\%)\\
        GPU ONN + $\widehat{\sin}$ & 99.95 (+0.5\%) & 90.60 (+9.1\%) & 68.40 (+9.5\%)\\ 
        \makecell{GPU ONN + $\widehat{\sin}$ +$\widehat{W_{i,j}}$} & 99.93 (+0.4\%) & 86.00 (+3.5\%) & 65.23 (+4.5\%)\\ 
        ODEONN & 99.50 & 83.08 & 62.45 \\ \hline
        \multicolumn{4}{|c|}{$7\times6$ Patterns} \\ \hline
        GPU ONN & 99.98 (+0.2\%) & 91.58 (+9.7\%) & 81.0 (+17\%) \\ 
        \makecell{GPU ONN +$\widehat{W_{i,j}}$} & 98.80 (-1.0\%) & 81.14 (-2.8\%) & 68.86 (-0.4\%) \\
        GPU ONN + $\widehat{\sin}$ & 100 (+0.2\%) & 94.56 (+13\%) & 82.28 (+19\%) \\ 
        \makecell{GPU ONN + $\widehat{\sin}$ +$\widehat{W_{i,j}}$} & 98.60 (-1.2\%) & 79.24 (-5.1\%) & 67.10 (-3.0\%) \\ 
        ODEONN & 99.76 & 83.50 & 69.14 \\ \hline
        \multicolumn{4}{|c|}{$10\times10$ Patterns} \\ \hline
        GPU ONN & 100 (+11\%) & 96.96 (+36\%) & 82.48 (+25\%) \\ 
        \makecell{GPU ONN +$\widehat{W_{i,j}}$} & 82.40 (-8.7\%) & 69.10 (-3.3\%) & 63.64 (-3.8\%) \\
        GPU ONN + $\widehat{\sin}$ & 99.98 (+11\%) & 97.74 (+37\%) & 81.02 (+22\%)\\ 
        \makecell{GPU ONN + $\widehat{\sin}$ +$\widehat{W_{i,j}}$} & 80.92 (-10\%) & 66.80 (-6.5\%) & 62.38 (-5.7\%)\\ 
        ODEONN & 90.24 & 71.46 & 66.18 \\ \hline
        \multicolumn{4}{|c|}{$22\times22$ Patterns} \\ \hline
        GPU ONN & 100 (+25\%) & 99.70 (+45\%) & 90.02 (+47\%) \\ 
        \makecell{GPU ONN +$\widehat{W_{i,j}}$} & 66.18 (-17\%) & 61.88 (-9.7\%) & 57.86 (-5.3\%) \\
        GPU ONN + $\widehat{\sin}$ & 100 (+25\%) & 100 (+46\%) & 88.60 (+45\%)\\ 
        \makecell{GPU ONN + $\widehat{\sin}$ +$\widehat{W_{i,j}}$} & 74.72 (-6.4\%) & 63.58 (-7.2\%) & 59.30 (-2.9\%)\\ 
        ODEONN & 79.78 & 68.54 & 61.08 \\ \hline
    \end{tabular}
    \label{tab:appendix:associative_memory}
    \begin{tablenotes}
        \item \textit{Note: The values in brackets are the relative difference to ODEONN.}
    \end{tablenotes}
\end{table}

\begin{table}
    \centering
    \captionsetup{width=\textwidth}
    \caption{Resource usage Excluding AXI interfaces}
    \label{tab:appendix:resource_usage}
    \begin{tabular}{|c|c|c|c|c|}
        \hline
        $N_\text{cores}$ & 16 & 32 & 64 & 128\tnote{a}      \\ \hline
        \multicolumn{5}{|c|}{$N_\text{banks}=1$}    \\
        \multicolumn{1}{|c}{$N_\text{oscillators}$} & \multicolumn{1}{c}{16} & \multicolumn{1}{c}{32} & \multicolumn{1}{c}{64} & \multicolumn{1}{c|}{128} \\ \hline
        LUTs    & 2146  & 4280  & 8732  & 25606     \\
        FFs     & 1875  & 3672  & 7278  & 17649     \\
        DSPs    & 32    & 64    & 128   & 220       \\
        BRAMs   & 8     & 16    & 32    & 64        \\ \hline
        \multicolumn{5}{|c|}{$N_\text{banks}=2$}    \\
        \multicolumn{1}{|c}{$N_\text{oscillators}$} & \multicolumn{1}{c}{32} & \multicolumn{1}{c}{64} & \multicolumn{1}{c}{128} & \multicolumn{1}{c|}{256} \\ \hline
        LUTs    & 2577  & 5081  & 10409 & 28908     \\
        FFs     & 2385  & 4696  & 9365  & 21768     \\
        DSPs    & 32    & 64    & 128   & 220       \\
        BRAMs   & 8     & 16    & 32    & 64        \\ \hline
        \multicolumn{5}{|c|}{$N_\text{banks}=4$}    \\
        \multicolumn{1}{|c}{$N_\text{oscillators}$} & \multicolumn{1}{c}{64} & \multicolumn{1}{c}{128} & \multicolumn{1}{c}{256} & \multicolumn{1}{c|}{512} \\ \hline
        LUTs    & 3539  & 7000  & 14434 & 36853     \\
        FFs     & 3423  & 6771  & 13543 & 29960     \\
        DSPs    & 32    & 64    & 128   & 220       \\
        BRAMs   & 8     & 16    & 32    & 128        \\ \hline
        \multicolumn{5}{|c|}{$N_\text{banks}=8$}    \\
        \multicolumn{1}{|c}{$N_\text{oscillators}$} & \multicolumn{1}{c}{128} & \multicolumn{1}{c}{256} & \multicolumn{1}{c}{512} & \multicolumn{1}{c|}{n/a\tnote{b}} \\ \hline
        LUTs    & 5137  & 10554 & 20261 & n/a       \\
        FFs     & 5493  & 10918 & 22687 & n/a       \\
        DSPs    & 32    & 64    & 128   & n/a       \\
        BRAMs   & 8     & 32    & 128   & n/a       \\ \hline
    \end{tabular}
    \begin{tablenotes}
        \item [a] At this point, the maximum number of DSPs provided by the FPGA is exceeded, the remaining logic has been absorbed into LUTs and FFs.
        \item [b] For $N_\text{banks} = 8$ and $N_\text{cores}=128$, the architecture is not synthesizable due to exceeding the limitations in memory resources.
    \end{tablenotes}
\end{table}

\end{appendices}

\end{document}